\documentstyle[psfig,11pt]{article}
\def\gsimeq
{\hbox{\raise0.5ex\hbox{$>\lower1.06ex\hbox{$\kern-1.07em{\sim}$}$}}}
\def\lsimeq
{\hbox{\raise0.5ex\hbox{$<\lower1.06ex\hbox{$\kern-1.07em{\sim}$}$}}}

\def\section#1{\par{#1}\par}

\begin{document}

%\doublespace

\vspace{1.0cm}
{\Large \bf DIFFUSE THERMAL EMISSION FROM VERY HOT GAS IN STARBURST GALAXIES:
 SPATIAL RESULTS}
%\footnote{Contributed talk to appear in {\it The 32$^{nd}$ COSPAR Meeting, The AGN-Galaxy Connection} (Advances in Space Research).}
%, ed. H.~R. Schmitt, A.~L. Kinney, and L.~C. Ho 
\vspace{1.0cm}

	M.~Cappi$^{1*}$,
        M.~Persic$^2$,
        S.~Mariani$^3$,
        L.~Bassani$^1$,
        L.~Danese$^4$,
        A.J.~Dean$^5$,
        G.~Di~Cocco$^1$,
        A.~Franceschini$^6$,
        L.K.~Hunt$^7$,
        F.~Matteucci$^8$,
        S.~Molendi$^9$
	E.~Palazzi$^1$,
        G.G.C.~Palumbo$^{2,1}$,
        Y.~Rephaeli$^{10}$,
        P.~Salucci$^4$, and 
        A.~Spizzichino$^1$

\vspace{1.0cm}
$^1$ ITeSRE/CNR, via Gobetti 101, 40129 Bologna, Italy 
\\
$^2$ Trieste Astronomical Observatory, via G.B.Tiepolo 11, 
        34131 Trieste, Italy   
\\
$^3$Astronomy Dept., University of Bologna, via Zamboni 33, 
        40126 Bologna, Italy    
\\
$^4$SISSA/ISAS, via Beirut 2-4, 34013 Trieste, Italy        
\\
$^5$Physics Dept., Southampton University, Southampton SO9 5NH, UK 
\\
$^6$Astronomy Dept., University of Padova, vicolo dell'Osservatorio 5, 
        35122 Padova, Italy                        
\\
$^7$CAISMI/CNR, Largo E.Fermi 5, 50125 Firenze, Italy 
\\
$^8$Astronomy Dept., University of Trieste, via Besenghi, 
        34131 Trieste, Italy                              
\\
$^9$IFCTR/CNR, Via Bassini 15, 20133, Milano, Italy
\\
$^{10}$School of Physics and Astronomy, Tel Aviv University, Tel Aviv 69978,
        Israel\\
\\
\\
\hspace{1truecm}{$^*$ e-mail: mcappi@tesre.bo.cnr.it}

\vspace{0.5cm}

%%%%%%%%%%%%%%%%%%%%
\section{ABSTRACT}
%%%%%%%%%%%%%%%%%%%%

New $BeppoSAX$ observations of the nearby prototypical starburst 
galaxies NGC 253 and M82 are presented. 
A companion paper (Cappi et al. 1998) shows that the hard (2-10 keV) spectrum 
of both galaxies, extracted from the source central regions, is best described 
by a thermal emission model with kT $\sim$ 6--9 keV and abundances 
$\sim$ 0.1--0.3 solar. The spatial analysis yields clear evidence that 
this emission is extended in NGC 253, and possibly also in M82. 
This quite clearly rules out a LLAGN as the main responsible for their hard 
X-ray emission.
Significant contribution from point-sources (i.e. X-ray binaries (XRBs) and 
Supernovae Remnants (SNRs)) cannot be excluded; neither can we at present 
reliably estimate the level of Compton emission.
However, we argue that such contributions shouldn't affect our main 
conclusion, i.e.,
that the $BeppoSAX$ results show, altogether, compelling evidence for 
the existence of a very hot, metal-poor interstellar plasma in both galaxies. 

%New $BeppoSAX$ observations of the nearby archetypical starburst 
%galaxies NGC 253 and M82 are presented. 
%The main observational result 
%is the first and unambiguous evidence that the hard (2-10 keV) component of both 
%galaxies is produced primarily by thermal emission from a metal-poor 
%($\sim$ 0.1--0.3 solar), hot (kT $\sim$ 6-- 9 keV) and extended plasma.
%Here we present only the results of the spatial analysis, briefly discuss 
%possible origins of the hot gaseous component, and assess its immediate 
%implications.
%Spectral results obtained for the 
%source central regions can be found in a companion paper (Cappi et al. 1998).

%%%%%%%%%%%%%%%%%%%%%%%
\section{PREVIOUS SPATIAL RESULTS}
%%%%%%%%%%%%%%%%%%%%%%%

Evidence for complex galactic-scale outflows driven by starburst activity has been 
gathered in recent years based, primarily, on optical and soft ($\sim$ 0.1--3 keV)
X-ray observations (Fabbiano 1989 and ref. therein). These have sometimes 
been called ``superbubbles'' or ``superwinds'', the latter referring 
to those manifestations where the extended hot gas emitting optical emission 
lines and soft X-rays was apparently ejected into the intergalactic medium (IGM).
Because of the extinction of the optical emission (often almost completely 
reprocessed into IR emission), X-ray data provided the most direct 
view of the hot wind material.   
In general, spectroscopic studies confirmed or, at least, were consistent with 
thermal emission from a hot plasma, most likely shock-heated by supernovae 
(e.g. Dahlem, Weaver \& Heckman 1998).
The detailed physical characteristics of the gas (in particular its metal 
abundance), however, have remained unclear because the analysis in the soft 
X-ray band is complicated by the unknown line-of-sight extinction, the large 
uncertainties in theoretical models used (in particular around the Fe L-shell 
energy band), and the presence of multiple temperatures (typically with kT between 
0.2--3 keV).

At higher energies, the images available to date have been 
essentially limited to the $ASCA$ observations of the 
two brightest starburst galaxies (SBGs), NGC 253 and M82 (but see recent 
studies on star-forming dwarf galaxies by Della Ceca et al. 1996, 1997). 
$ASCA$ resolved the 2--10 keV emission of NGC 253 (Ptak et al. 1997) 
but not from M 82 (Tsuru et al. 1997).
However, the {\it ASCA} data did not allow adequate spatial analysis.
From the {\it ASCA} spectral analysis, the hard components 
of both sources are well described by either a thermal model (kT $\sim$ 6--9 keV), 
or a power-law model ($\Gamma$ $\sim$ 1.8--2.0). 
The absence in the data of a significant Fe-K line emission 
has, however, always been puzzling and has induced several authors to propose 
alternative explanations to the thermal emission, i.e. the presence of a low-luminosity 
active galactic nucleus (LLAGN) (Ptak et al. 1997, Tsuru et al. 1997), or 
non-thermal emission from Compton scattering of relativistic electrons by the 
intense FIR radiation field (Rephaeli et al. 1991, Moran \& Lehnert 1997).

In a companion paper (Cappi et al. 1998), we have presented the $BeppoSAX$ spectral 
results that clearly show the first evidence of Fe-K line emission (at $\sim$ 6.7 keV) 
and high-energy rollover expected in the case of thermal emission for both 
NGC 253 and M82 (see Persic et al. 1998 for more details on the results for NGC 253).
Here we present preliminary results obtained from the spatial analysis which 
support the thermal origin of the hard component in NGC 253 and, to a 
lesser extent, in M82.

\section{$BEPPOSAX$ IMAGES OF NGC 253 AND M82}

$BeppoSAX$ observed NGC 253 on Nov.29--Dec.2, 1996 
and M82 on Dec.06--07, 1997 with the LECS, MECS and PDS detectors operating 
between 0.1--4 keV, 1.3--10 keV and 13--60 keV, respectively (see Table 1).
The spectral results have demonstrated that two thermal components (kT $\sim$ 
0.1--0.3 keV and kT $\sim$ 6--9 keV) are required to fit the spectra of both 
sources (Cappi et al. 1998, Persic et al. 1998), and that the hard thermal component 
needs to be absorbed in order not to over-produce the continuum at E $\lsimeq$1 keV.
Therefore, the present analysis will focus on the spatial 
properties in only the 3--10 keV energy range, where the contribution of the soft 
component is marginal.
Here we present preliminary results obtained from only the MECS instruments.

\begin{table}[hbt]
\begin{center}   
{\bf Table1:} Exposure times and count-rates \\
\begin{tabular}{cccc}
\hline
Source &  Inst. & Exposure & Count-rate \\
       & & Ksec & 10$^{-2}$ cts/s \\
\hline
NGC 253 & LECS & 55 & 3.9 \\
 	& MECS & 113 & 9.2 \\
 	& PDS(13-100 keV) & 51 & 7 ($\sim$ 2.5$\sigma$ detection) \\
& & & \\
M82	& LECS & 29 & 19 \\
 	& MECS & 85 & 35 \\
 	& PDS(13-25 keV) & 30 & 8 ($\sim$ 6$\sigma$ detection)\\
\hline
\end{tabular}
\end{center}
\end{table}

\begin{figure}[htb]
\parbox{8truecm}
{\psfig{file=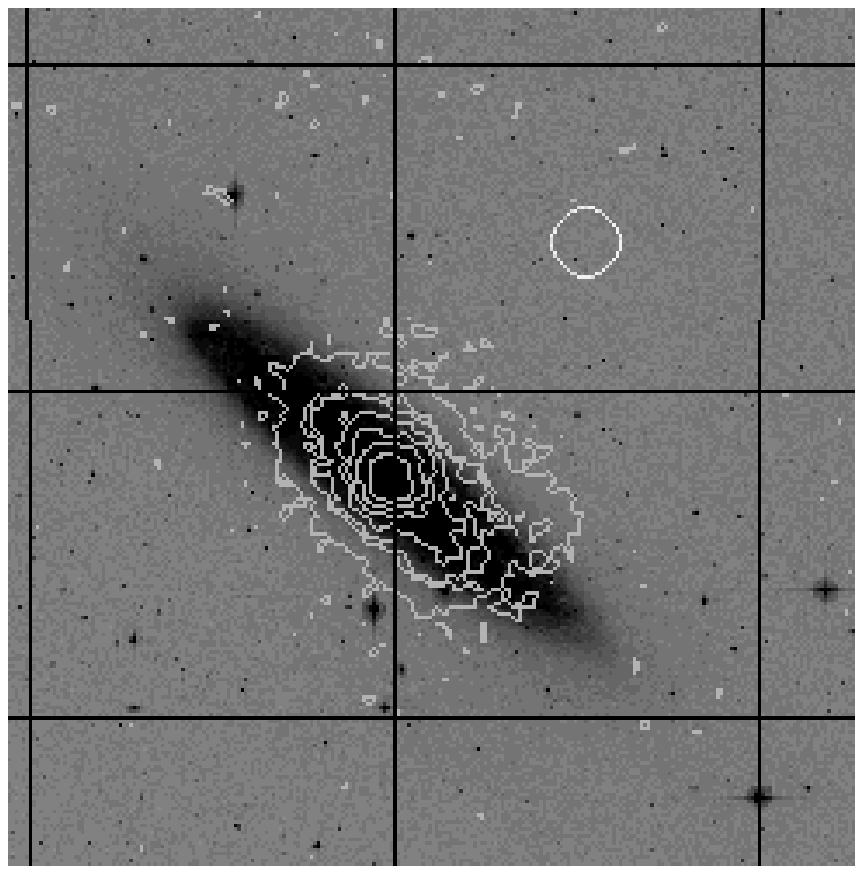,width=7cm,height=7cm,angle=0}}
\ \hspace{3truecm} \
\parbox{8truecm}
{\psfig{file=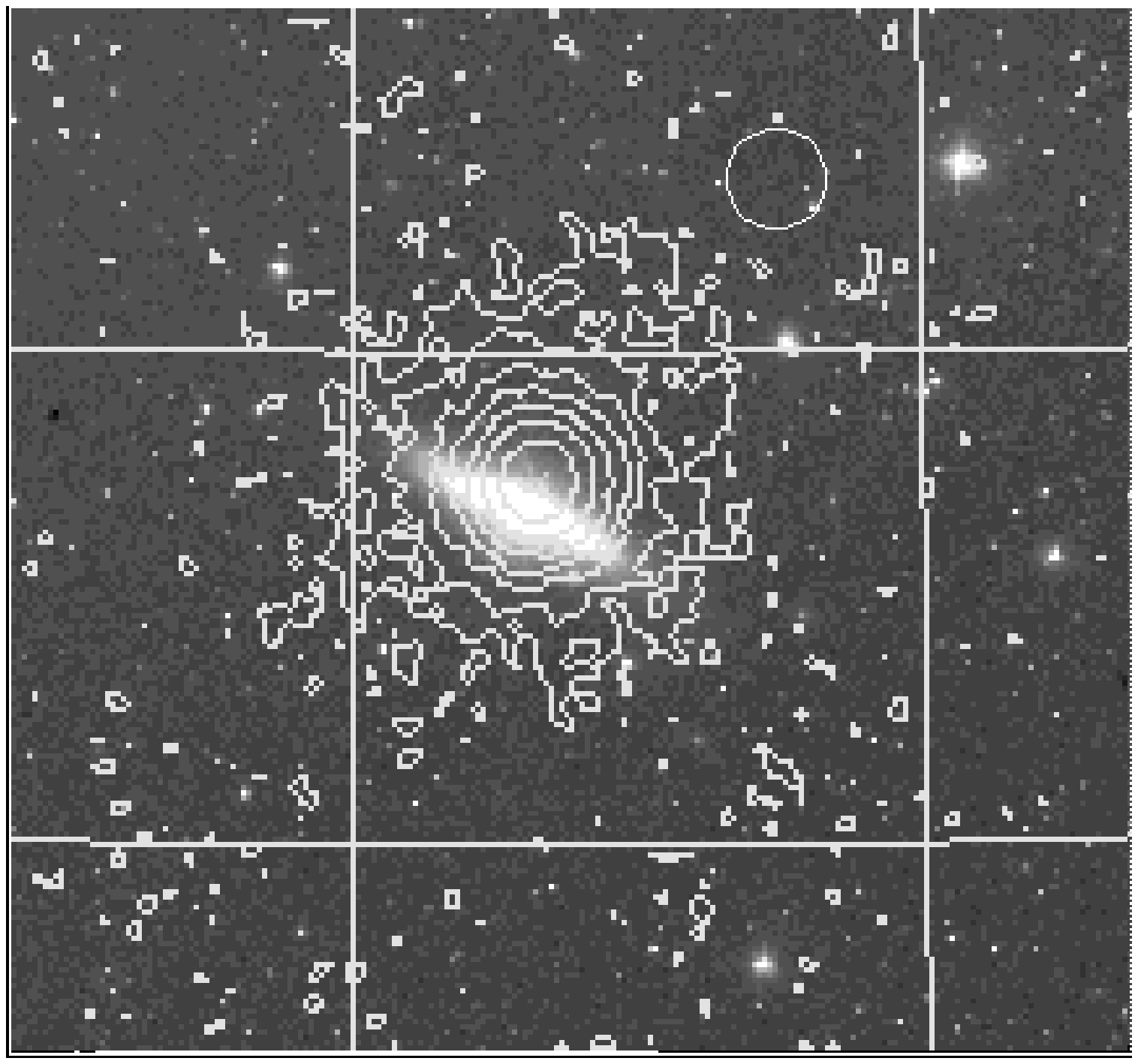,width=7cm,height=7cm,angle=0}}
\\
\\
{{\bf Fig.1:} $BeppoSAX$ MECS 3--10 keV image of NGC 253 (left) and M82 (right), 
superimposed on DSS images. The displayed FOV are $\sim$ 38$^{\prime}$ 
$\times$ 33$^{\prime}$ for NGC 253 and $\sim$ 25$^{\prime}$ 
$\times$ 25$^{\prime}$ for M 82. The circle on the top-right of each source 
represents the MECS 50\% Half Power Radii averaged over all energies (radius 
$\sim$ 1.3$^{\prime}$). The apparent shift between the centroid of the X-ray contours 
and of the DSS image of M82 is within the systematics (of $\sim$ 1$^{\prime}$) in the 
absolute position determination of $BeppoSAX$.}
\end{figure}

Figure 1 shows the MECS 3--10 keV images of both galaxies superimposed on  
Digital Sky Survey images. 
The left panel clearly shows that the hard X-ray emission of NGC 253 is extended 
and elongated along its major axis. No point sources embedded in the extended emission 
are detected, but given the limited resolution they cannot be ruled out; note that there 
is an indication for a ``cone'' of X-ray emission that extends toward the 
southwest direction. Another interesting feature is the apparent 
extended emission perpendicular to the major axis in the northwest and southeast, 
in a way similar to the ROSAT PSPC and HRI results (Dahlem, Weaver \& Heckman 1998). 
As a matter of fact, such emission could be the signature of a very hot 
gas ejected out of the galaxy, into the IGM.
Further analysis work is in progress to determine the detailed characteristics 
(e.g. flux, temperature vs. distance) of this extended component.
The right panel suggests the presence of a more symmetric X-ray halo in M82, though with 
some excess emission oriented along the optical minor axis in the northwest direction.

The radial profiles of the 3--5 keV, 7--10 keV and 6--7 keV emission from 
NGC 253 and M82 are shown in Figure 2 together with the instrumental PSF energy-weigthed 
over the source spectra.
The new and most convincing result of the present analysis 
is that the hard X-ray emission in NGC 253 
extends to $\sim$ 8$^{\prime}$. There is also evidence that M82 extends 
to $\sim$ 5$^{\prime}$, however the effect is in this case only marginal.
In NGC 253, the extension is also evident if one considers the FeK line flux 
only (i.e. in the 6--7 keV band).

\begin{figure}[htb]
\vspace{1truecm}
\parbox{8truecm}
{\psfig{file=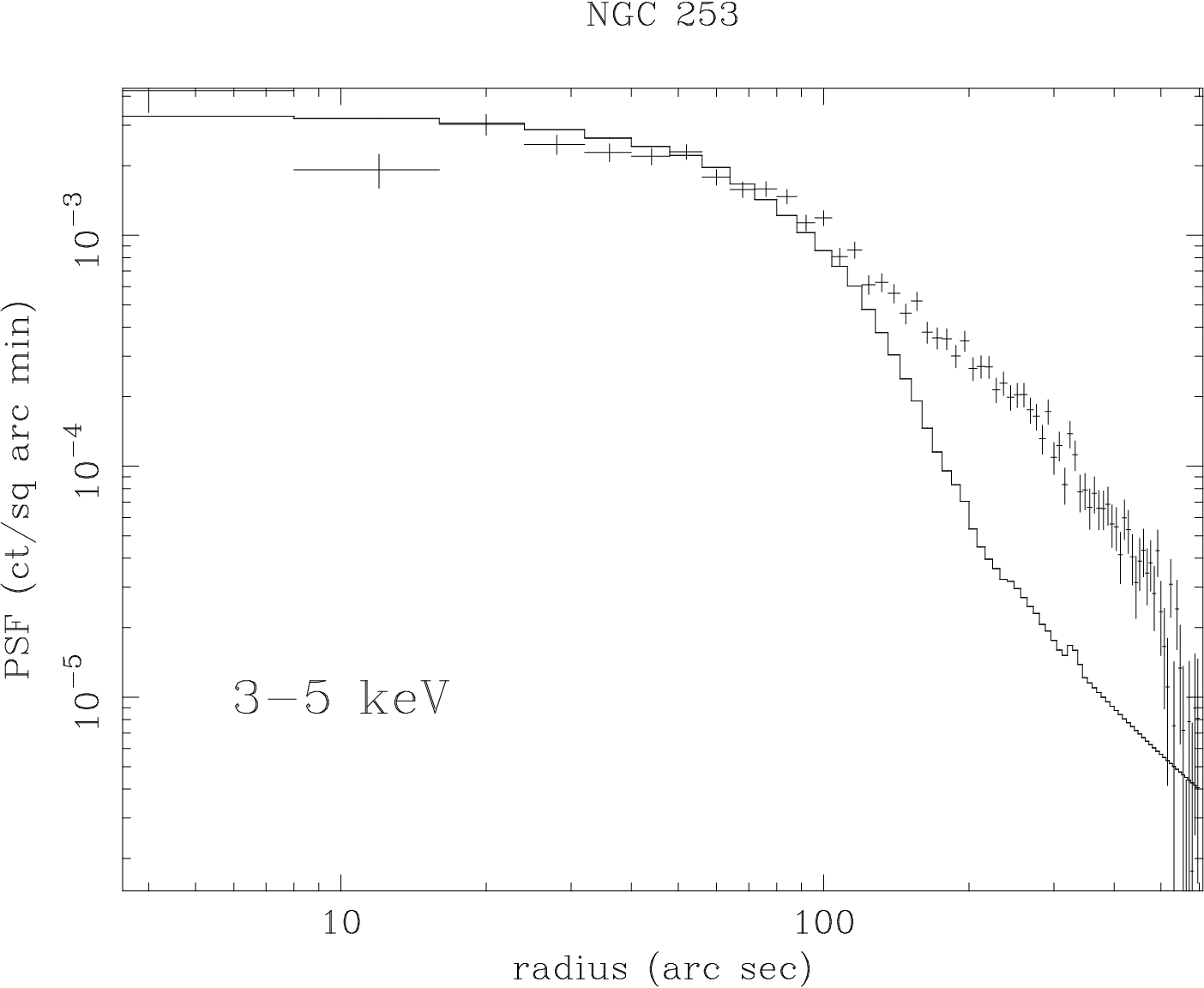,width=8.5cm,height=2.5cm,angle=-90}}
\ \hspace{0.5truecm} \
\parbox{8truecm}
{\psfig{file=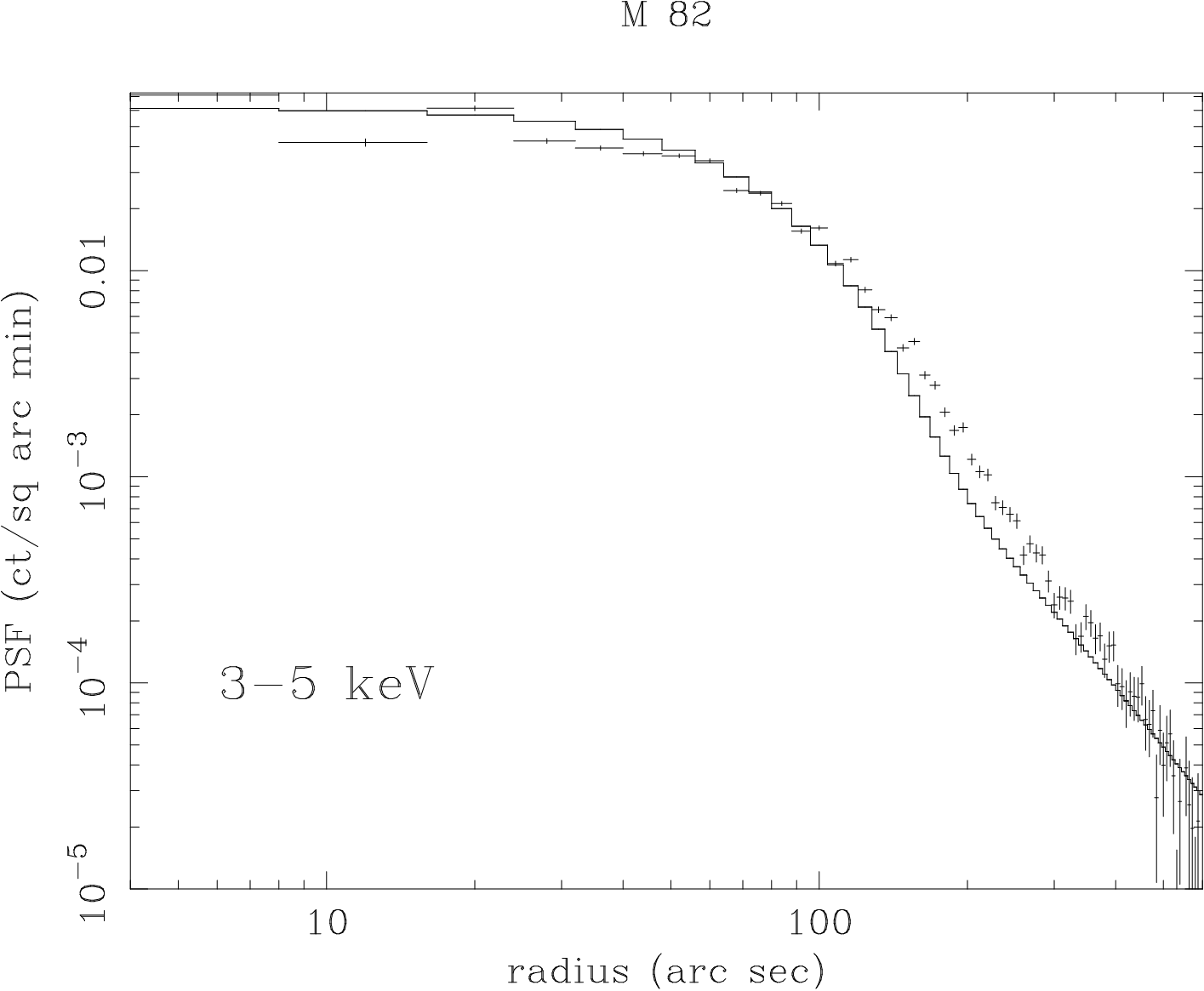,width=8.5cm,height=2.5cm,angle=-90}}
\parbox{8truecm}
{\psfig{file=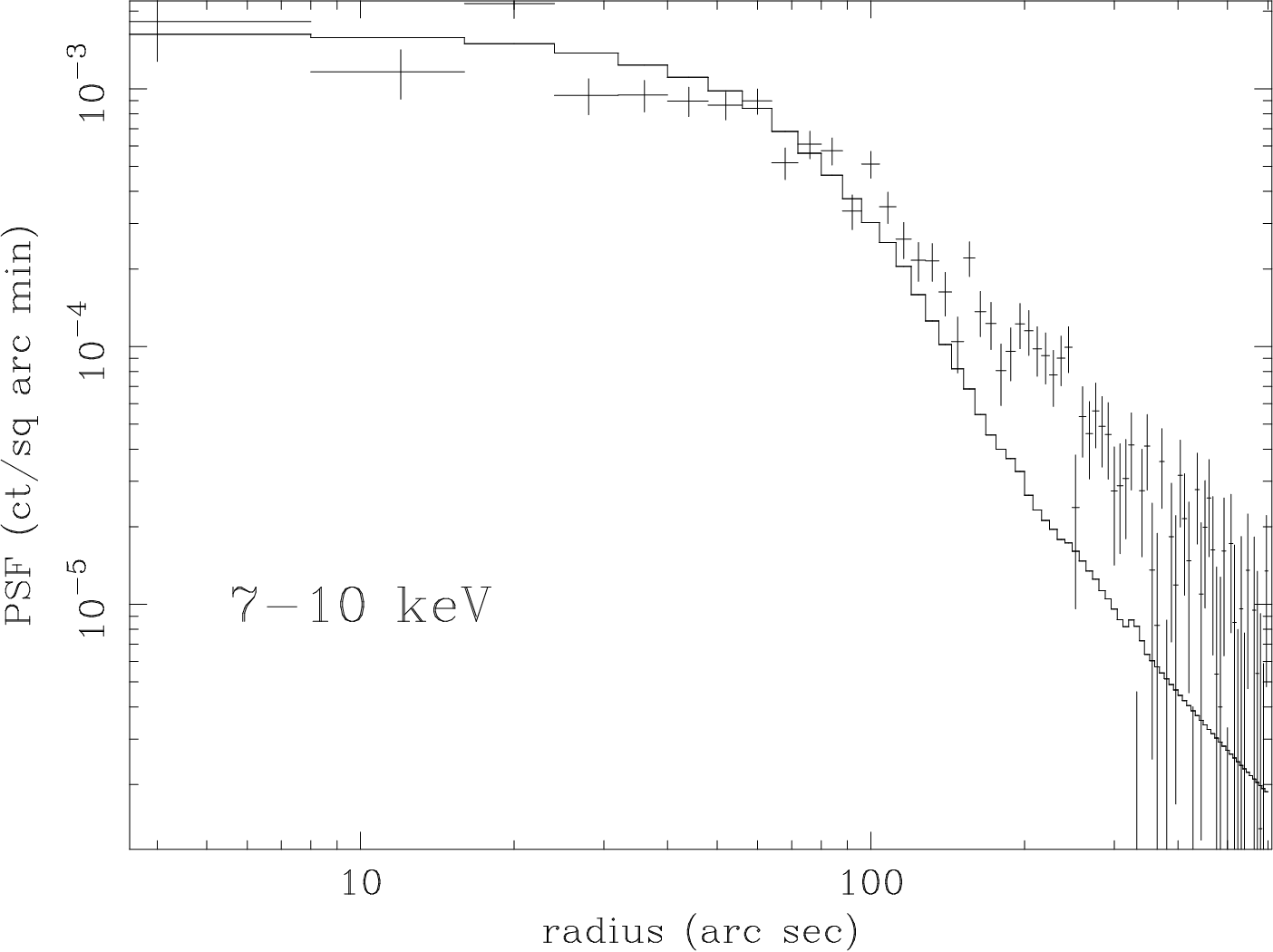,width=8.5cm,height=2.5cm,angle=-90}}
\ \hspace{0.5truecm} \
\parbox{8truecm}
{\psfig{file=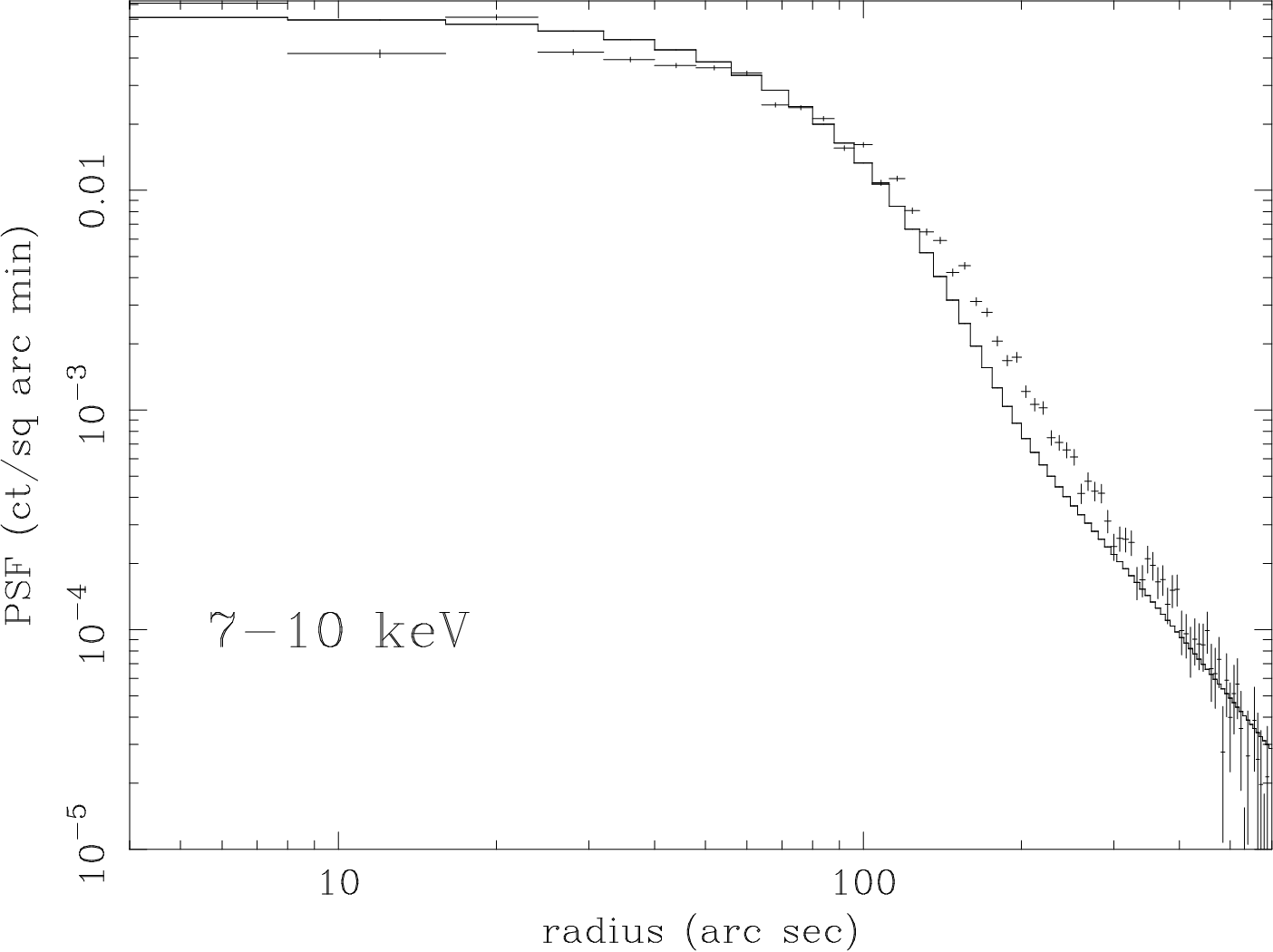,width=8.5cm,height=2.5cm,angle=-90}}
\parbox{8truecm}
{\psfig{file=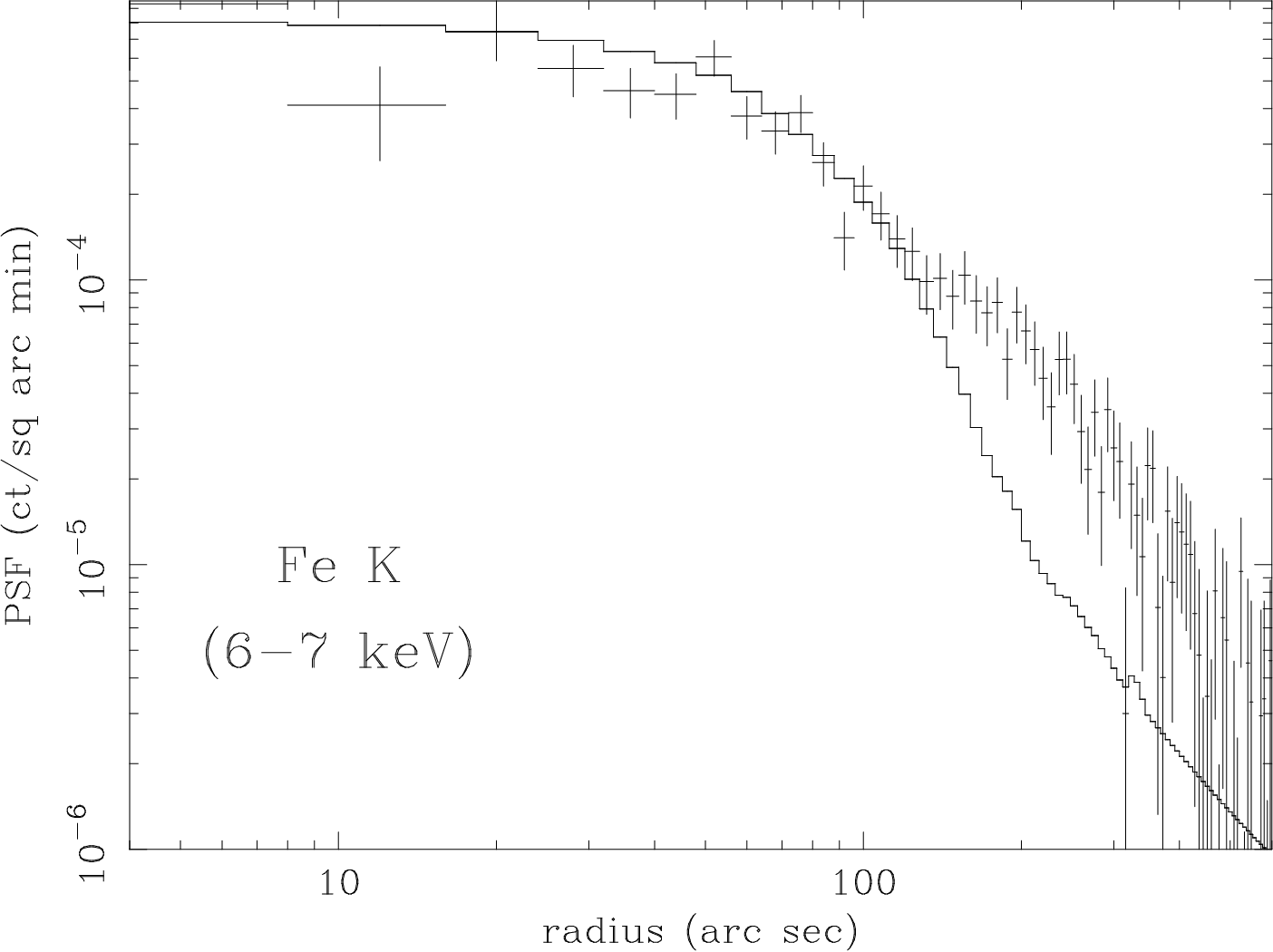,width=8.5cm,height=2.5cm,angle=-90}}
\ \hspace{0.5truecm} \
\parbox{8truecm}
{\psfig{file=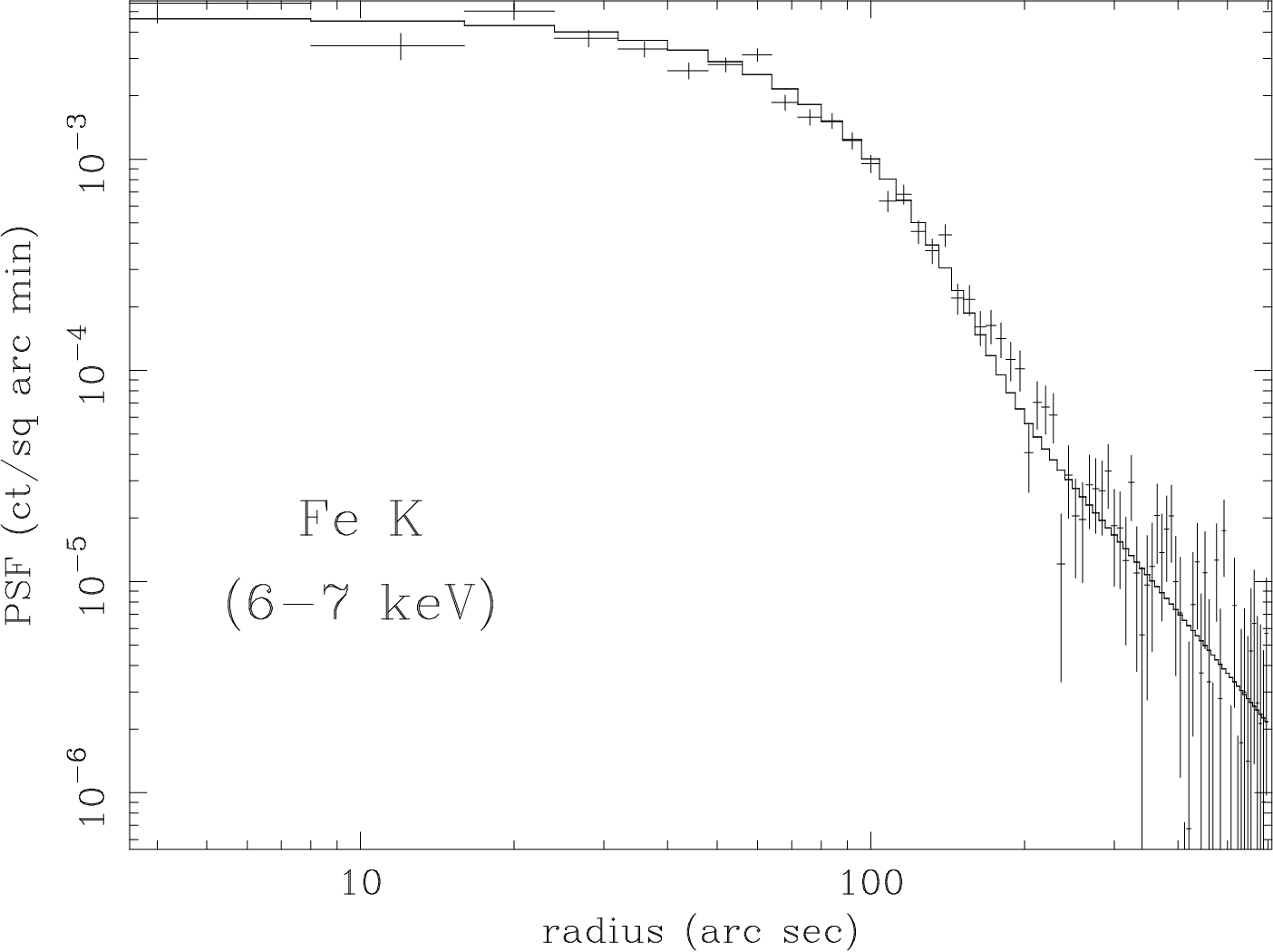,width=8.5cm,height=2.5cm,angle=-90}}
\vspace{0.4truecm}
\\
\\
{{\bf Fig.2:} Radial profiles (data points) of the sources emission between 3--5 keV 
(upper pannels), 7--10 keV (middle pannels) and 6--7 keV (lower pannels) as compared 
to the PSF for an on-axis point source (solid line) calculated over the 
corresponding energy range weighting the different energies with the sources spectra. 
Extended emission is clearly detected (for all energy bands) in NGC 253 and 
marginally in M82.}
\end{figure}

%Preliminary results obtained from the spatially-resolved spectral analysis 
%of NGC 253 are shown in Figure 3. Spectral fits were performed between 3--10 keV 
%in 9 concentric annuli centered on the source and with similar statistics.
%They suggest a decrease of the temperature as the distance from the source 
%increases (with a significance $>$ 99.95\%).

\section{ON THE ORIGIN OF THE HARD SPECTRAL COMPONENT}

The origin of the hard extended component is puzzling. It could be due to either 
a collection of point sources which contribute to the 
3-10 keV emission (e.g., XRBs, SNRs), or a truly diffuse emission 
due to Compton scattering of IR-optical photons from relativistic e$^{-}$. Alternatively, it 
could be a very hot ($\sim$ 6--10 $\times$ 10$^7$ K) ISM phase the cause of the hard 
extended component.
{\it Individually}, each of these components, except for the hot ISM phase hypothesis, 
seems unlikely to {\it dominate} the 3-10 keV of these galaxies.
The average spectrum of an ensemble of SNRs would probably be too soft 
(kT$\lsimeq$ 4 keV), and Compton emission would predict a spectrum with a power-law shape 
at odds with what shown in the companion paper (Cappi et al. 1998).
An accurate study by Dahlem, Weaver \& Heckman (1998) based on spatially resolved ROSAT 
PSPC spectra of NGC 253 and M82, has shown that the 0.1-2 keV flux of 
NGC 253/M82 can be divided into emission from the source disk+core (53\%/82\%), 
the halo (25\%/11\%) and point sources (22\%/7\%). 
Extrapolating the average spectrum of the ROSAT point sources of 
NGC 253 and M82 to higher energies, we obtain a 3-10 keV flux of 
$\sim$ 1.4 $\times$ 10$^{-12}$ 
erg cm$^{-2}$s$^{-1}$ and 5 $\times$ 10$^{-14}$ erg cm$^{-2}$s$^{-1}$, respectively.
This is negligible in the case of M82 (F$_{3-10 keV}$ $\sim$ 2.3 $\times$ 10$^{-11}$ 
erg cm$^{-2}$s$^{-1}$), and less than 40\% of the flux of NGC 253 (F$_{3-10 keV}$ $\sim$ 
3.6 $\times$ 10$^{-12}$ erg cm$^{-2}$s$^{-1}$). 
In conclusion, alternatives to the hot ISM phase could hardly produce, by themselves, 
all the hard X-ray emission detected in both galaxies. Therefore the interpretation 
according to which most of the hard X-ray emission is produced in a hot ISM plasma 
seems to be favored.

However, as mentioned above for NGC 253, XRBs (and possibly Compton emission, Rephaeli 
et al. 1991) certainly contribute to the hard X-ray emission.
Thus we estimated how their contribution would modify our conclusions on the 
measurements of temperature and abundance of the hard component. To do so, we added 
an extra hard component (an absorbed power-law) to the best-fit spectra shown 
in Cappi et al. (1998) to mimic the extra contribution from XRBs and/or 
Compton emission and/or emission from a LLAGN. 
For several values of $N_{\rm H}$ (from 10$^{22}$ to 10$^{24}$ cm$^{-2}$) 
and $\Gamma$ (from 1 to 2), we found no improvement of the fit and upper-limits 
of (at most) 20\% and 10\% of the observed 3-10 keV flux of NGC 253 and M82, 
respectively. In NGC 253, forcing the power-law contribution to be about 40\% of the 
total 3-10 keV flux, the thermal component softens from kT $\sim$ 6 keV to $\sim$ 4.8 keV and 
the abundances increase from $\sim$ 0.25 to $\sim$ 0.34 solar.
In M82, forcing a power-law contribution of 50\% of the total 3-10 keV flux, the 
thermal component softens from kT $\sim$ 8 keV to $\sim$ 6 keV, and the abundances 
increase from 
$\sim$ 0.08 to $\sim$ 0.25 solar, but the fit becomes worse by $\Delta \chi^2$ = 11 
(mainly because of the clear cutoff obtained from the PDS data). 
However, it should be pointed out that the average spectrum of XRBs 
may not be well described by a single absorbed power-law but might require the 
addition of an FeK line emission (most XRBs are known to emit strong FeK lines 
at 6.4 and/or 6.7 keV). In such case, abundance differences would become even lower.

It should be noted also that in the case of M82, some short-term ($\sim$ hrs) variability 
(with $\sim$ 30\% amplitude) was 
detected in the 3--10 keV light curve of M82, possibly indicating a contribution 
from XRBs to the hard X-ray flux. Given the lack of strong point-sources 
in the ROSAT PSPC observations of M82 reported by Dahlem, Weaver \& Heckman (1998), these 
could either be highly-variable (as suggested by Ptak et al. 1997), or be strongly 
absorbed in order to show up at E$\gsimeq$ 3 keV. However, our timing analysis indicates 
a dispersion of the light curve around its mean value of only (15$\pm$ 4)\%.
Therefore, as shown above, such a contribution should not have strong effects on 
our spectral results.

The overall results presented here are thus consistent with a {\it major} contribution 
in the 3-10 keV band from a hot and diffuse thermal plasma. 
A significant contribution from other emission 
mechanisms (point-source population and/or Compton emission) cannot be excluded based on the 
present data but even with the extreme hypothesis of a $\sim$ 30\% contribution between 
3-10 keV, best-fit temperatures and abundances derived from the hard component would become 
only slightly lower and higher than reported.
In any case, the present results clearly rule out the possibility that a LLAGN makes the 
bulk of the hard X-ray emission in these two SBGs. 

The above is consistent with our preliminary results obtained from a spatially-resolved 
spectral analysis of the MECS data of NGC 253 (Cappi et al., in prep.) 
which clearly shows that the 
temperature of the hard component decreases with increasing distance from the core 
(from kT $\sim$ 6 keV in the core to kT $\sim$ 4 keV in the disk). 
This suggests either a thermal contribution from point-source populations 
in the core and in the disk (but both with thermal average spectra) or thermal emission 
from a hot ISM gas.

%Given the too limited spatial resolution of BeppoSAX, we cannot exclude that all of 
%these components do contribute (at least in part) to the hard X-ray emission. 
%However, if one uses a ``zero-order'' approximation that {\it all} the 3-10 keV 
%emission shall be produced by a {\it single} component

If due to hot gas, the observed temperatures (T$_{\rm obs.}$ 
$\sim$ 6.5/9.7 $\times$ 10$^7$ K for NGC253/M82) are much higher than 
the ``escape temperature'' (T$_{\rm esc.}$ $\sim$ 2/1 $\times$ 10$^6$ K for 
NGC253/M82; Wang et al. 1995) of the gas in these 
galaxies, so the gas should easily escape from the galaxies.
As pointed out by Heckman (1997), this could consequently be very important 
for understanding galactic evolution and the chemical enrichment 
of IC and IG gas.

%Moreover, as suggested in Cappi et al. (1998), there is a striking 
%similiraty between the $BeppoSAX$ spectra of NGC 253 and M82 with 
%the $ASCA$ spectrum of the Galactic Ridge. Moreover, as 

Finally, it is interesting to note the analogy of our results with those 
on clusters of galaxies obtained about 25 years ago.
Indeed, the Perseus, Virgo and Coma clusters were known to be X-ray sources but the 
origin of their 2--10 keV emission was at first unclear.
Several hypotheses were proposed: e.g., a collection of 
AGNs or AGNs-related (point) sources (Kellog et 1972), Compton scattering 
of microwave background photons 
by electrons emitting the diffuse radio halos (Forman et al. 1972) or a hypothetical 
intracluster medium (Gott and Gunn 1971). But it was only after the first observational 
evidence with $UHURU$ that cluster 2--10 keV emission was shown to be extended 
(Forman et al. 1972, Kellog et al. 1975) {\it and} by the detection by $Ariel-5$ 
(Mitchell 1976) and $OSO$-8 (Serlemitsos et al. 1977) of an iron emission line at 6.7 keV
that the origin of their hard X-ray emission was attributed to 
a hot, evolved, and diffuse intracluster gas. Does history repeat itself ?

\section{CONCLUSIONS}

The main conclusion from the above results is that the $BeppoSAX$ observations 
of the two prototypical SBGs NGC 253 and M82 have revealed for the first
time evidence of {\it a very hot and diffused thermal plasma} which is mainly responsible 
for their hard (2--10 keV) emission. This discovery is likely to have important 
implications on the AGN/starburst connection and on our understanding 
of the chemical enrichment of the IG medium and on the formation and evolution of galaxies.

\section{REFERENCES}
\vspace{-5mm}
\begin{itemize}
\setlength{\itemindent}{-8mm}
\setlength{\itemsep}{-1mm}

\item [] Cappi, M., {\it et al.}, to appear in proceedings of ``Dal nano- al tera-eV: tutti i 
colori degli AGN'', third Italian conference on AGNs, Roma, {\it Memorie S.A.It}, astro-ph/9809325 (1998)

\item [] Dahlem, W., Weaver, K.A., \& Heckman, T.M., {\it Astrophys. J. Suppl.}, in press (1998)

\item [] Della Ceca, R., Griffiths, R.E., \& Heckman, T.H., {\it Astrophys. J.}, {\bf 485}, 581 (1997)

\item [] Della Ceca, R., Griffiths, R.E., Heckman, T.H., \& MacKenty, J.W., {\it Astrophys. J.}, 
{\bf 469}, 662  (1996)

\item [] Fabbiano, G., $A.R.A.\&A.$, {\bf 27}, 87 (1989)

%\item [] Heckman, T.M, To appear in the proceedings of the conference `The Most Distant Radio Galaxies', D. Reidel (astro-ph/9801155)

\item [] Forman, W., Kellogg, E., Gursky, H., Tananbaum, H., \& Giacconi, {\it Astrophys. J.}, {\bf 178}, 309 (1972)

\item [] Gott, J., \& Gunn, J., {\it Astrophys. J.}, {\bf 169}, L13 (1971)

\item [] Heckman, T.M, {\it RevMexAA (Seie de Conferencias)}, {\bf 6}, 156 (1997)

\item [] Kellog, E., Baldwin, J.R., \& Koch, D., {\it Astrophys. J.}, {\bf 199}, 299 (1975)

\item [] Kellog, E., Gursky, H., Tananbaum, H., \& Giacconi, R., {\it Astrophys. J.}, {\bf 174}, L65 (1972)

\item [] Mitchell, R.J., Culhane, J.L., Davison, P.J.N., \& Ives, J.C., {\it M.N.R.A.S.}, {\bf 176}, 29 (1976)

\item [] Moran, E.C, \& Lehnert, D., {\it Astrophys. J.}, {\bf 478}, 172 (1997)

\item [] Persic, M., {\it et al.}, {\it Astron. Astr.}, {\bf 339}, L33 (1998)

\item [] Ptak, A., Serlemitsos, P., Yaqoob, T., Mushotzky R. \& Tsuru, T., 
{\it Astron. J.}, {\bf 113}, 1286 (1997)

\item [] Rephaeli, Y., Gruber, D., Persic, M., \& D. McDonald, {\it Astrophys. J.}, {\bf 380}, L59 (1991)

\item [] Serlemitsos, P.J., Smith, B.W., Boldt, E.A., Holt, S.S., \& Swank, 
J.H., {\it Astrophys. J.}, {\bf 211}, L63 (1977)

\item [] Tsuru, T.G., Awaki, H., Koyama, K., \& Ptak, A., {\it P.A.S.J.}, in press (1998)

\item [] Wang, D., Walterbos, R., Steakley, M., Norman, C., \& Braun, R., {\it Astrophys. J.}, {\bf 439}, 176 (1995)

\end{itemize}

\end{document}